\documentclass[aps,prd,reprint]{revtex4-1}

\usepackage{url}
\usepackage{natbib}
\usepackage{textcase}
\usepackage[colorlinks=true,
            linkcolor=blue,
            citecolor=blue]{hyperref}

\usepackage[ngerman, english]{babel}
\usepackage[utf8]{inputenc}
\usepackage{amsmath, amssymb}
\usepackage{slashed}		
\usepackage{braket} 

\usepackage{tabularx} 
\usepackage{colortbl} 
\usepackage{subfigure}
\usepackage{bbold} 

\usepackage{placeins} 
\usepackage{paralist} 

  \usepackage{graphicx}
  \usepackage{epstopdf}	

\def\nn{\nonumber}

\begin{document}
\title{Direct- and Resolved-Photon Threshold Resummation for Polarized High-$p_T$ Hadron Production at COMPASS}
\author{Claudia Uebler}
\author{Andreas Schäfer}
\affiliation{Institute for Theoretical Physics, University of Regensburg, D-93053 Regensburg, Germany}
\author{Werner Vogelsang}
\affiliation{Institute for Theoretical Physics, Tübingen University, D-72076 Tübingen, Germany}
\begin{abstract}
We complete our earlier study of the ``direct'' part of the cross section and spin asymmetry for the photoproduction 
process $\gamma N\rightarrow h X$ 
by analysing the ``resolved'' contribution, for which the photon couples like a hadron through 
its parton structure. The incident photon and nucleon are
longitudinally polarized and one observes a hadron $h$ at high transverse momentum $p_T$.
Soft or collinear gluon emissions generate large logarithmic threshold corrections which we 
resum to next-to-leading logarithmic order.
We compare our results with recent spin asymmetry data by the COMPASS collaboration,
highlighting the role of the fragmentation functions.
\end{abstract}

\maketitle
\section{Introduction}
The question of how the gluon and quark spins and orbital angular momenta
combine to generate the nucleon spin of 1/2 has been under much debate for 
a long time. Experimental information comes from high-energy scattering processes
involving longitudinally polarized nucleons, among them semi-inclusive hadron photoproduction 
$\gamma N \rightarrow h X$ which is used by the COMPASS experiment at CERN
to shed new light on nucleon spin structure. 
COMPASS has presented results for the spin-averaged cross section already a while ago~\cite{Adolph:2012nm},
and more recently also for the corresponding  
double-longitudinal spin asymmetry $A_{\mathrm{LL}}$~\cite{Adolph:2015hta,Levillain:2015twa}. The latter is directly 
sensitive to the spin-dependent gluon distribution $\Delta g$, which in turn provides information on 
the gluon spin contribution to the proton spin. Although they still have rather limited precision, the COMPASS data are 
complementary to the probes of $\Delta g$ employed at RHIC~\cite{Lin:2017gzi}. 

Reliable information on $\Delta g$ may only be obtained from the data if the theoretical 
framework is adequate for describing $\gamma N \rightarrow h X$ in the kinematic
regime relevant at COMPASS. While hard photoproduction is in principle well 
understood, in particular the relation between the  ``direct'' and ``resolved'' contributions (see, 
for example, Ref.~\cite{Klasen:2002xb}), it has been pointed out~\cite{MelaniesPaper} that there 
are large QCD corrections for COMPASS kinematics that require resummation to all orders.
Basically, at COMPASS one is relatively close to a kinematic threshold that arises when 
nearly all available energy of the incoming partons is used for the production of the high-$p_T$ 
final state and its recoiling counterpart. The phase space for radiation of additional partons then 
becomes small, resulting in large logarithmic ``threshold'' corrections at every order in perturbation
theory from the cancelation of infrared  divergences between real and virtual diagrams.
These threshold logarithms may be resummed to all 
orders of perturbation 
theory~\cite{Sterman1987,Catani:1989ne,LaenenJuni1998,KidonakisJan1998,KidonakisMa1998,Bonciani:2003nt,
Catani:2013vaa,Becher:2009th}.
In Ref.~\cite{MelaniesPaper} we have performed such a threshold resummation at
next-to-leading logarithmic (NLL) accuracy for the spin-averaged cross section
for $\gamma N \rightarrow h X$, finding that the resummed cross section
shows a markedly better agreement with the experimental data than the fixed-order (next-to-leading
order, NLO) one. In our recent paper~\cite{Uebler:2015ria} we have extended our calculations to the 
case of longitudinally polarized incoming photons and nucleons, considering first the direct part
of the cross section for which the photon interacts in the usual point-like manner in the hard scattering. 
In the present paper we complete our study of the polarized case by also addressing the resolved-photon 
contributions where the photon reveals its partonic structure and interacts like a hadron. 
Again we present studies that incorporate all-order QCD threshold resummation and thereby 
improve the theoretical framework relevant for comparison to the COMPASS data. 

Our paper is organized as follows. In Section~\ref{sec2} we briefly review the general framework for 
photoproduction cross sections in perturbative QCD. In Section~\ref{sec3} we recall the threshold resummation 
formalism for (resolved) photoproduction. The relevant techniques are rather standard, and numerous details
have been given in our previous papers~\cite{MelaniesPaper,Uebler:2015ria}, so we shall be brief here
as well. Section~\ref{sec4} presents phenomenological results for our theoretical predictions for the 
spin-dependent cross sections, as well as a comparison of our double-longitudinal spin asymmetries with 
COMPASS data. Finally, we summarize and conclude in Section~\ref{sec5}. 

\section{Photoproduction cross section in perturbation theory \label{sec2}}

\begin{figure}[t]
      \centering
\vspace*{-1.3cm}
      \includegraphics[width=0.75\linewidth]{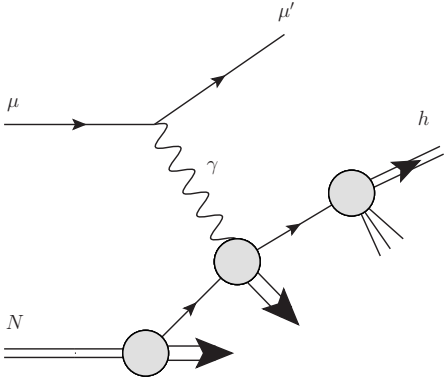}
\vspace*{-1.5cm}
      \caption{\label{lepton-nucleon-dir} {\it Single-inclusive 
high-$p_T$ hadron production in lepton scattering (direct contribution).}}
\end{figure}

We consider the process (see Fig. \ref{lepton-nucleon-dir})
\begin{align}
\ell N \rightarrow \ell' h X\,,
\end{align}
in which the initial lepton $\ell$ scatters off a nucleon $N$, both longitudinally polarized,
and (semi-inclusively) produces a charged hadron $h$ with high transverse momentum $p_T$. 
The scattered lepton $\ell'$ is demanded to have a small scattering angle with respect to the initial one, 
so that the underlying process can be treated as a {\it photoproduction} process
$\gamma N \rightarrow h X$, for which the main contributions come from almost on-shell photons exchanged
between the lepton and the nucleon.

At sufficiently large transverse momentum of the observed hadron, perturbative-QCD techniques may be applied. 
The differential spin-dependent cross section $d\Delta \sigma$ as function of $p_T$ and the hadron's
pseudorapidity $\eta$ may be written in factorized form as~\cite{Jaeger2003,Jaeger2005}:
\begin{align}
	& \!\!\!\! \frac{p_T^3 d \Delta \sigma}{dp_T d\eta}= \sum_{abc} \int_{x_\ell^{{\mathrm{min}}}}^{1}  \!\!\!\! dx_\ell 
	 \int_{x_n^{{\mathrm{min}}}}^{1} \!\!\!\! dx_n \int_{x}^{1} \!\! dz   \nonumber \\[2mm]
	& \times\frac{\hat{x}_T^4 z^2 }{8v} 
	 \frac{\hat{s} d\Delta\hat{\sigma}_{ab\rightarrow cX}(v,w,\hat{s},\mu_R,\mu_F,\mu_F')}{dv \, dw}  \nonumber \\[2mm]
	 &\times \Delta f_{a/\ell} \! \left( x_\ell, \mu_F \right) \Delta f_{b/N} \! \left( x_n, \mu_F \right)  D_{h/c} \! \left( z, \mu_F' \right) \,,
	 \label{ordinary_factorization}
\end{align}
where the sum runs over all possible partonic channels $a b \rightarrow c X$. 
$\Delta f_{a/\ell} \left( x_{\ell}, \mu_F \right)$ and $\Delta f_{b/N} \left( x_n, \mu_F \right)$
denote the polarized distribution functions for partons $a$ and $b$ in the lepton and the nucleon, respectively,
depending on the momentum fractions $x_{\ell}$ and $x_n$ and on the initial-state factorization scale $\mu_F$. 
The $D_{h/c}(z, \mu_F' )$ are the parton-to-hadron fragmentation functions describing the hadronization of parton
$c$. They depend on the fraction $z$ of the parton's momentum taken by the hadron and on the final-state factorization
scale $\mu_F'$. Finally, the $d\Delta\hat{\sigma}_{ab\rightarrow cX}$ are the differential spin-dependent 
partonic hard-scattering cross sections, which are perturbative and thus can be expanded in terms of the strong 
coupling constant $\alpha_s$:
\begin{align}
d\Delta\hat{\sigma}_{ab\rightarrow cX} = d\Delta\hat{\sigma}_{ab\rightarrow cX}^{(0)} + 
\frac{\alpha_s}{\pi}d\Delta\hat{\sigma}_{ab\rightarrow cX}^{(1)} + ... \, .
\end{align}
They depend on the factorization scales $\mu_F$, $\mu_F'$, on the renormalization scale $\mu_R$ and 
on the kinematic variables introduced in Eq.~(\ref{ordinary_factorization}):
\begin{align}\label{vwdef}
 v\equiv 1 +\frac{\hat{t}}{\hat{s}} \quad \text{and} \quad w \equiv \frac{- \hat{u}}{\hat{s}+ \hat{t}}\,,
\end{align}
with the partonic Mandelstam variables $\hat{s}$, $\hat{t}$ and $\hat{u}$. Moreover,
\begin{align}
\hat{x}_T\equiv \frac{x_T}{z \sqrt{x_\ell x_n}} \quad \text{and} \quad 
\hat{\eta}\equiv \eta - \frac{1}{2} \ln \frac{x_\ell}{x_n}\,,
\end{align}
where $x_T\equiv 2 p_T/\sqrt{S}$, with $\sqrt{S}$ and $\eta$ being the hadronic center-of-mass energy and
rapidity, respectively, the latter counted positive in the lepton forward direction. 
The lower integration bounds in Eq. (\ref{ordinary_factorization}) are given by
\begin{align}
 x_\ell^{{\mathrm{min}}}&= \frac{x_T e^{\eta} }{2 - x_T e^{- \eta}}, \nonumber \\[2mm]
 x_n^{{\mathrm{min}}} & = \frac{x_\ell x_T e^{-\eta} }{2 x_\ell - x_T e^{\eta}}, \nonumber \\[2mm]
x&=\frac{x_T \cosh \hat{\eta}}{\sqrt{x_n x_\ell}}\,.
\end{align}
We note that all formulas presented so far equally apply to the spin averaged case by simply 
dropping the ``$\Delta$'' everywhere, so that the unpolarized partonic cross sections and parton distributions 
appear. 

\begin{figure}[t]
      \centering
      \vspace*{-1.3cm}
      \includegraphics[width=0.75\linewidth]{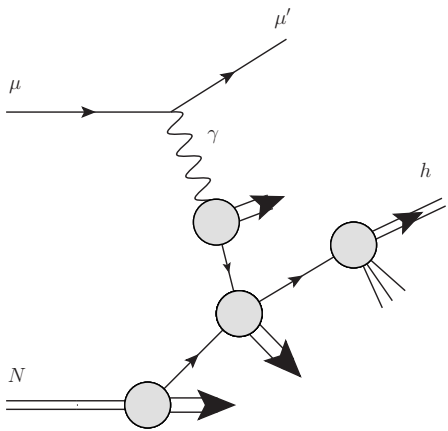}
\vspace*{-1.3cm}
      \caption{\label{lepton-nucleon-resolved} {\it Resolved-photon contribution to high-$p_T$ hadron
      production in lepton scattering.}}
\end{figure}

As is well known, the physical photoproduction cross section is the sum of two parts: 
\begin{align}
  d \Delta \sigma =  d \Delta \sigma_{{\mathrm{dir}}} +  d \Delta \sigma_{{\mathrm{res}}}\,,
\end{align}
where for the {\it direct} part $d \Delta \sigma_{{\mathrm{dir}}}$ the photon couples directly in a point-like way  
to the parton $b$ in the nucleon, while for $d \Delta \sigma_{{\mathrm{res}}}$ it couples 
to quantum fluctuations containing quarks, antiquarks and gluons and hence
is resolved into its own partonic structure. For a quasi-real photon such contributions are not suppressed 
by additional, strongly virtual propagators, and the physical photon eigenstate contains an appreciable QCD part. 
This part is described by photonic parton distribution functions just as for normal hadrons, see  
Fig. \ref{lepton-nucleon-resolved}. The resulting contribution is called \textit{resolved} photon contribution.

One can accommodate both the direct and the resolved contributions by introducing suitable
``parton-in-lepton'' distributions~\cite{Jaeger2003,Jaeger2005}: 
\begin{align}
  \Delta f_{a/\ell} (x_\ell, \mu_F) = \int_{x_\ell}^{1} \frac{dy}{y} 
  \Delta P_{\gamma \ell}(y) \Delta f_{a/ \gamma} \left(x_\gamma = \frac{x_\ell}{y}, \mu_F\right) \,,
  \label{PDF_lepton}
\end{align}
which are convolutions of the probability density $\Delta P_{\gamma \ell } (y)$ to have a polarized 
(``Weizs\"{a}cker-Williams'') photon with lepton momentum fraction $y$ accompanying the 
initial-state lepton, and the probability density $\Delta f_{a/ \gamma}(x_\gamma,\mu_F)$ to find a 
polarized parton $a$ with momentum fraction $x_{\gamma}$ in this photon. In case
of the direct contributions, one simply has $\Delta f_{\gamma / \gamma}= \delta \left( 1 - x_{\gamma}\right)$.
$\Delta P_{\gamma \ell } (y)$ is given by~\cite{deFlorian:1999ge}
\begin{align}
 \Delta P_{\gamma \ell} (y) =& \frac{\alpha}{2\pi} 
 \left[ \frac{1 - \left(1 - y \right)^2 }{y} \ln{ \left( \frac{ Q_{{\mathrm{max}}}^2 (1-y)}{m_\ell^2 y^2} \right)} \right. 
 \nonumber \\[2mm]
  &  \hspace{6mm} \left.  + 2 m_\ell^2  y^2 \left( \frac{1}{Q_{{\mathrm{max}}}^2} - \frac{1-y}{m_\ell^2 y^2} \right) \right] \,.
 \label{WeizsaeckerSpectrum}
\end{align}
Here $\alpha$ is the fine structure constant, $m_\ell$ the lepton mass, 
and $Q_{{\mathrm{max}}}^2$ the maximum 
value of virtuality $Q^2$ allowed by the experimental conditions on the small-angle scattered lepton. 

The direct contribution starts at lowest order (LO) with the subprocesses $\gamma q\to qg$ and $\gamma g
\to q\bar{q}$, which are of order ${\cal O}(\alpha\alpha_s)$. For the resolved contributions, the lowest
order partonic processes are the $2\to 2$ QCD ones
\begin{align}\label{qcd}
  &q\,q' \rightarrow q \, q', \quad  q \, \bar{q}' \rightarrow q  \,\bar{q}', \quad  q \, \bar{q}  
  \rightarrow q' \, \bar{q}', \quad   q \,q \rightarrow q \,q , \nonumber \\
  & q \, \bar{q} \rightarrow q \bar{q} , \quad q \, \bar{q} \rightarrow g \,g, \quad  g \, 
  q \rightarrow q  \,g, \quad  g \, g \rightarrow g \,q, \nonumber \\
  & g \,g \rightarrow g \,g , \quad  g \,g \rightarrow q  \,\bar{q} .
\end{align}
These are of order ${\cal O}(\alpha_s^2)$. Since, however, the photon's parton distributions 
$\Delta f_{a/ \gamma}$ are of order $\alpha/ \alpha_s$~\cite{Klasen:2002xb}, the resolved 
contribution is of the same perturbative order as the direct one. This remains true to all orders.
NLO (${\cal O}(\alpha\alpha_s^2)$) QCD corrections to polarized high-$p_T$ photoproduction 
of hadrons have been derived in~\cite{deFlorian1998,Jaeger2003,Jaeger2005}.
At LO, the partonic cross sections are always proportional to $\delta(1-w)$ since $(1-w)$ measures
the invariant mass of the partonic recoil. At NLO various types of distributions in $(1-w)$ arise. 
Analytical expressions may be found in 
Refs.~\cite{deFlorian1998,Jaeger2003,Jaeger2002,Aversa:1988vb,Gordon:1994wu,Hinderer:2015hra}.
They can be cast into the form
\begin{align}
 \frac{\hat{s} d \Delta \hat{\sigma}^{(1)}_{a b\rightarrow c X} (v,w) }{dv \, dw}
 =& A(v) \delta(1-w) + B(v) \left( \frac{\ln(1-w)}{1-w} \right)_{+} \nonumber \\
  + &C(v) \left( \frac{1}{1-w} \right)_{+} + F(v,w) \text{,}
 \label{NLO_shaped}
\end{align}
where the plus-distributions are defined through
\begin{align}
 \int^1_0  dw \, f(w) \left[ g(w) \right]_{+} \equiv  \int^1_0  dw \,\left[  f(w) - f(1) \right] g(w) .
\end{align}
The functions $A(v),B(v),C(v),F(v,w)$ in~(\ref{NLO_shaped}) depend on the partonic process under consideration.
$F(v,w)$ collects all terms without distributions in $(1-w)$. 
The terms with plus-distributions give rise to the large double-logarithmic threshold corrections
that are addressed by resummation. They arise from soft-gluon radiation and 
recur with higher logarithmic power at every higher order of $\alpha_s$. At the 
$k$-th order we have leading corrections proportional to $\alpha_s^k [\ln^{2k-1} (1-w) / (1-w)]_{+}$;
subleading terms are down by one or more powers of $\ln(1-w)$.
We next describe the all-order resummation of these threshold logarithms.

\section{Resummed cross section \label{sec3}}
Our goal is to resum the logarithmic corrections to next-to-leading logarithmic (NLL) level, 
so that we take into account the three most leading ``towers'' $\alpha_s^k  [\ln^{2k-1} (1-w) / (1-w)]_{+}$, 
$\alpha_s^k  [\ln^{2k-2} (1-w) / (1-w)]_{+}$ and  $\alpha_s^k  [\ln^{2k-3} (1-w) / (1-w)]_{+}$ to all orders of
perturbation theory. This is also what we have previously achieved for the spin-averaged cross
section~\cite{MelaniesPaper} and in our study~\cite{Uebler:2015ria} of the direct part of the spin-dependent
cross section. Here we extend the resummation to the polarized resolved part. We note that the cross section for
the resolved part of $\gamma N \rightarrow h X$ has the same structure as that for,
say, $pp\to hX$, so that many of the relevant techniques are rather well-established in the 
literature~\cite{LaenenJuni1998,KidonakisJan1998,KidonakisMa1998,Bonciani:2003nt,Catani:2013vaa},
where spin-averaged cross sections were considered. We also note that in Ref.~\cite{deFlorian2007} threshold 
resummation for polarized hadronic scattering $pp\to hX$ was investigated where, however, only the cross section
integrated over all rapidities of the produced hadron was considered. Here we perform the resummation
at {\it arbitrary fixed} rapidity of the produced hadron, using the techniques developed 
in Refs.~\cite{Sterman2000,Almeida2009,Hinderer:2014qta,MelaniesPaper,Uebler:2015ria}.

\subsection{Transformation to Mellin moment space}
We perform the resummation in Mellin moment space. Starting from
Eq.~(\ref{ordinary_factorization}) we write the convolution of the partonic cross section with the fragmentation
function as the Mellin inverse of the corresponding product of Mellin moments. In this way we find
(see~\cite{Almeida2009,MelaniesPaper}):
\begin{align}
 	 \!\! \frac{p_T^3 d\sigma}{dp_T d\eta}= &\sum_{abc} \int_{0}^{1}  \!\! dx_\ell \int_{0}^{1} \!\! dx_n 
	 \Delta f_{a/\ell}  \left( x_\ell, \mu_F \right) \Delta f_{b/N}  \left( x_n, \mu_F \right) \nonumber \\[2mm]
	 \times & \int_{\mathcal{C}} \frac{dN}{2 \pi i} (x^2)^{-N} D_{h/c}^{2N+3} (\mu_F') 
	 \Delta \tilde{w}_{a b\rightarrow cX}^{2N} \left( \hat{ \eta} \right),
	 \label{factorization}
\end{align}
where $D_{h/c}^N (\mu) \equiv \int_0^1 dz \, z^{N-1} D_{h/c} (z, \mu)$. 
The integration contour $\mathcal{C}$ is initially a line from $C-{\rm i}\infty$ to 
$C+{\rm i}\infty$, with the positive real number $C$ chosen such that  $\mathcal{C}$ 
passes to the right of all singularities of the integrand. In numerical applications the
contour is usually tilted with respect to the real axis as described in~\cite{MelaniesPaper,Uebler:2015ria}, in order 
to improve numerical convergence of the Mellin integral.
The hard-scattering function in Mellin moment space is given by 
\begin{align}
 \Delta \tilde{w}_{a b\rightarrow cX}^{N} (\hat{\eta}) \equiv 
 2 \! \! \int^1_0 d \zeta \left(\! 1 -\zeta  \right)^{ \! N-1}  \! \frac{\hat{x}_T^4 z^2}{8v} 
 \frac{\hat{s}d\Delta\hat{\sigma}_{a b\rightarrow cX}}{dv \, dw} \text{,}
 \label{hard_scattering_Mellin}
\end{align}
with
$\zeta\equiv v (1-w) = 1-\hat{x}_T \cosh \hat{\eta}$.
Note that  $\Delta \tilde{w}_{a b\rightarrow cX}^N$ depends on $\hat{s}$ and on the factorization and renormalization scales.
In Mellin space, the logarithms $\alpha_s^k  [\ln^m (1-w) / (1-w)]_{+}$ discussed above turn into logarithms
of the form $\alpha_s^k \ln^{m+1}(N)$.
As seen from Eq.~(\ref{factorization}), we keep the parton distribution functions in $x$-space:
The moments of the fragmentation functions in~(\ref{factorization}) alone lead to a sufficiently fast fall-off
of the integrand of the inverse Mellin transform 
that the convolution with the parton distribution functions can be carried out 
numerically. 

\subsection{NLL-resummed hard-scattering function}

To NLL accuracy, the resummed hard-scattering function reads (see
\cite{LaenenJuni1998,KidonakisJan1998,Sterman2000,MelaniesPaper}):
\begin{align}
 \Delta \tilde{w}_{a b \rightarrow cd}^{\text{resum}, N} \left(\hat{\eta} \right) 
      =& \Delta_a^{N_a}(\hat{s},\mu_F,\mu_R)
      \Delta_b^{N_b} (\hat{s},\mu_F,\mu_R) \nonumber \\[1mm]
      \times & \Delta_c^{N} (\hat{s},\mu_F',\mu_R) J_d^{N}(\hat{s},\mu_R) \nonumber \\[1mm]
      \times & 
      \text{Tr} \left\{ \Delta H \mathcal{S}^{\dagger}_N S \mathcal{S}_N \right\}_{ab\rightarrow cd} \, ,
      \label{general_w}
 \end{align}
with $N_a=(-\hat{u}/\hat{s})N$ and $N_b=(-\hat{t}/\hat{s})N$. 
The functions $\Delta_a^{(-\hat{u}/\hat{s})N}$, $\Delta_b^{(-\hat{t}/\hat{s})N}$ and $\Delta_c^{N}$
are spin-independent. They 
describe soft gluon radiation collinear to the initial-state parton $a$, the initial-state parton $b$  and  the 
fragmenting parton $c$, respectively. They are exponentials and given in the $\overline{{\mathrm{MS}}}$ scheme 
by~\cite{LaenenJuni1998}
\begin{align}
  &  \ln \Delta_i^{N}\left(\hat{s},\mu_F,\mu_R \right)=  \nonumber \\[2mm]
    &  - \int_0^1 dz \frac{z^{N-1}-1}{1-z} \int_{(1-z)^2}^1 \frac{dt}{t} A_i\left(\alpha_s(t \hat{s})   \right)  
    \nonumber \\[2mm]
    &  -2 \int_{\mu_R}^{\sqrt{\hat{s}}} \frac{d\mu'}{\mu'} \gamma_i (\alpha_s(\mu'^2))
    + 2 \int_{\mu_F}^{\sqrt{\hat{s}}} \frac{d\mu'}{\mu'} \gamma_{ii} (N, \alpha_s(\mu'^2)), 
\label{ln_Delta}
\end{align}
where the functions $A_i,  \gamma_i ,\gamma_{ii}$ ($i=q,g$) are perturbative series
in the strong coupling and are given explicitly for example in \cite{Uebler:2015ria}. 
The function $J_d^{N}$ describes soft and hard
collinear emission off the unobserved recoiling parton $d$. We have~\cite{LaenenJuni1998}
\begin{align}
&    \ln J_d^{N}(\hat{s},\mu_R) =\int_0^1 dz \frac{z^{N-1}-1}{1-z}  \nn\\[2mm]
&  \times \bigg\{ \int_{(1-z)^2}^{(1-z)}\frac{dt}{t}  A_d \left(\alpha_s(t \hat{s})\right) - \gamma_d \left(\alpha_s((1 \! -z)\hat{s} )\right)  
		\bigg\} \nonumber \\[2mm]
		&\! + 2 \int_{\mu_R}^{\sqrt{\hat{s}}} \frac{d\mu'}{\mu'} \gamma_d \left(\alpha_s(\mu'^2)\right).
	\label{ln_J}	
\end{align}
Each of the functions in the trace term in (\ref{general_w}) is a matrix in the space of color exchange operators, 
and the trace is taken also in this space \cite{Almeida2009,KidonakisMa1998,KidonakisJan1998}.
The function $\Delta H_{ab\rightarrow cd}$  describes the spin-dependent hard-scattering and $S_{ab\rightarrow cd}$ is a soft function
for wide-angle gluon radiation. Following the formalism of \cite{Almeida2009} one can expand each of the 
functions perturbatively, so that the hard-scattering function reads:
\begin{align}
 \Delta H_{ab\rightarrow cd} (\hat{\eta},\alpha_s ) =  \Delta H_{ab\rightarrow cd}^{(0)} (\hat{\eta} ) + \frac{\alpha_s}{\pi}  \Delta H_{ab\rightarrow cd}^{(1)} (\hat{\eta} ) 
 + \mathcal{O}(\alpha_s^2)).
\end{align}
The lowest-order terms are given in \cite{deFlorian2007}. Analogously we get for the soft function
\begin{align}
 S_{ab\rightarrow cd} (\hat{\eta},\alpha_s ) =  S_{ \! ab\rightarrow cd}^{(0)}  \! + \frac{\alpha_s}{\pi} 
 S_{ \! ab\rightarrow cd}^{(1)} \left(\! \!\hat{\eta},\alpha_s,\! \frac{\sqrt{\hat{s}}}{N} \right) 
 + \mathcal{O}(\alpha_s^2)).
\end{align}
In the latter the $N$ dependence shows up only at next-to-next-to-leading  logarithmic level (NNLL). The LO terms 
$S_{ab\rightarrow cd}^{(0)}$ are $\eta$-independent (and spin-independent) and may for example be found in \cite{Kidonakis2000gi}. Finally, the 
functions $\mathcal{S}_{N,ab \rightarrow cd}$ are path-ordered exponentials of integrals over soft anomalous dimension matrices 
$\Gamma_{ab\rightarrow cd}$ \cite{Almeida2009,KidonakisMa1998,KidonakisJan1998}:
\begin{align}
 \mathcal{S}_{N,ab \rightarrow cd} \left(\hat{\eta}, \alpha_s \right)=\mathcal{P} \exp 
 \left[ \int_{\mu_R}^{\sqrt{\hat{s}}/N} \frac{d \mu'}{\mu'}  \Gamma_{ab\rightarrow cd} (\hat{\eta}, \alpha_s(\mu') )  \right].
\end{align}
The soft anomalous dimension matrices start at $\mathcal{O}(\alpha_s)$,
\begin{align}
  \Gamma_{ab\rightarrow cd} (\hat{\eta}, \alpha_s )  = \frac{\alpha_s}{\pi} \Gamma_{ab\rightarrow cd}^{(1)} (\hat{\eta} ) + \mathcal{O}(\alpha_s^2),
\end{align}
and can be found to first order in \cite{KidonakisMa1998,KidonakisJan1998,Kidonakis2000gi,Sjodahl2009}. 
According to \cite{Almeida2009,Catani:2013vaa}, to NLL we can approximate
\begin{align}
\! \! \text{Tr} \left\{ \! \Delta H \mathcal{S}^{\dagger}_N S \mathcal{S}_N \! \right\}_{ab\rightarrow cd} \!\! 
\approx & (1 \!+ \!\frac{\alpha_s}{\pi} \Delta C^{(1)}_{ab\rightarrow cd} )\nonumber \\[1mm]
\times & \text{Tr}\left\{ \Delta H^{(0)} \mathcal{S}^{\dagger}_N S^{(0)} \mathcal{S}_N  \right\}_{ab\rightarrow cd} ,
\end{align}
where we have introduced $N$-independent, spin-dependent hard-scattering 
coefficients $ \Delta C^{(1)}_{ab\rightarrow cd}$ that are determined by the $\Delta H_{ab\rightarrow cd}^{(1)}$
and $S_{ \! ab\rightarrow cd}^{(1)}$. 
They originate from virtual corrections at NLO and match the resummed cross section to the NLO one. Hence we can extract
them by comparing the first-order expansion of the resummed partonic cross section with the exact NLO 
one from~\cite{Jaeger2002}. The coefficients for the direct case have been published in our previous paper \cite{Uebler:2015ria}.

We are now ready to insert all ingredients into Eq.~(\ref{general_w}) and expand the result to NLL. Such expansions
are standard; see, for example, Refs.~\cite{Uebler:2015ria,MelaniesPaper}, and we do not provide them here. 
We have checked that all single- and double-logarithmic terms of the exact NLO cross section are reproduced by 
our NLL partonic cross section.

\subsection{Inverse Mellin transform and matching procedure}
After carrying out the resummation procedure for the hard scattering, we have to perform an inverse Mellin transform
as seen in Eq.~(\ref{factorization}). Here we have to deal with singularities appearing in the NLL exponents 
caused by the Landau pole in the perturbative strong coupling. As in our earlier papers \cite{MelaniesPaper,Uebler:2015ria}
we use the {\it Minimal Prescription} method introduced in \cite{Catani_MinimalPrescr} and choose the 
integration contour such that the Landau poles lie to the right of the contour.  
Furthermore, in order to make sure that NLO is fully included in our theoretical predictions, we match our 
resummed cross section to the NLO one. 
For this we subtract the first-order contributions that are present in the resummed expression 
and add the full NLO cross section:
\begin{align}
    \frac{p_T^3 \Delta d \hat{\sigma}^{\text{matched}} }{dp_T \, d\eta}= & \frac{p_T^3 d \Delta \sigma^{\text{NLO} } }{dp_T \, d\eta}
   +\sum_{bc} \int_0^1 dx_\ell \int_0^1 dx_n 
   \nonumber \\[2mm]
    \times & \Delta  f_{\gamma/\ell}(x_\ell,\mu_F)\Delta f_{b/N}(x_n,\mu_F) \nonumber \\[2mm]
     \times & \int_{\mathcal{C}} \frac{dN}{2 \pi i} (x^2)^{-N} D_{h/c}^{2N+3} (\mu_F') \nonumber \\[2mm]
    \times & \left[ \Delta \tilde{w}_{\gamma b \rightarrow cd}^{2N,\text{resum}}(\hat{\eta})  -  \left. 
   \Delta \tilde{w}_{\gamma b \rightarrow cd}^{2N,\text{resum}}(\hat{\eta}) \right|_{\mathrm{NLO}} \,\right] ,   
   \label{final_matched}
\end{align}
where ``$|_{\mathrm{NLO}}$'' labels the truncation at NLO. 
Hence, without any double-counting of perturbative terms, we take into account 
the NLL soft-gluon corrections beyond NLO, as well as the
full available fixed-order cross section.
\section{Phenomenological results \label{sec4}}

The goal of our calculation is to provide improved theoretical predictions for the 
polarized cross section and the double-longitudinal spin asymmetry for 
the photoproduction process $\mu N \rightarrow \mu' h X$, and to compare with the published 
COMPASS data~\cite{Adolph:2012nm,Adolph:2015hta}. COMPASS uses a longitudinally polarized muon 
beam with a mean beam energy of $E_\mu=160$ GeV. With a stationary nucleon target 
(we approximate the deuteron as a system of a free proton and neutron) this corresponds to a center-of-mass 
energy of $\sqrt{S}=17.4$ GeV. COMPASS adopts  $Q_{{\mathrm{max}}}^2=1~{\rm GeV}^2$ for the 
maximal virtuality of the exchanged photon, which we use in the Weizs\"{a}cker-Williams 
spectrum~(\ref{WeizsaeckerSpectrum}). We also introduce the COMPASS cut $0.2\leq z \leq 0.8$
for the energy fraction of the virtual photon carried by the hadron, as well as their cut 
on the lepton's momentum fraction carried by the photon, $0.1 \leq y \leq 0.9$. 
(Note that in our previous publication~\cite{Uebler:2015ria} we used 
$0.2 \leq y \leq 0.9$; however the impact of the precise choice is small.)
The scattering angle of the detected hadrons lies between $10\leq \theta \leq 120$ mrad, corresponding 
to the pseudorapidity range $-0.1 \leq \eta \leq 2.38$. When analyzing the asymmetries, we consider this 
full rapidity range as well as the smaller bins $[-0.1,0.45]$, $[0.45,0.9]$ and $[0.9,2.4]$.

For the calculations of the unpolarized NLL resummed cross section we follow Ref.~\cite{MelaniesPaper} 
and use the numerical code of that work.
Unless stated otherwise, we choose the renormalization and factorization scales as the 
transverse momentum of the produced charged hadron, $\mu_R=\mu_F=\mu_F'\equiv \mu=p_T$. 
To have some confidence that our perturbative methods are valid, we require the hadron transverse 
momentum to be at least  $p_T$=1.75 GeV, although experimental data are available down to 
$p_T$=0.7 GeV.

We use the helicity dependent parton distributions of Ref.~\cite{deFlorian:2014yva} (``DSSV2014'')
and the unpolarized ones of Ref.~\cite{Martin:2009iq} (``MSTW''). For the resolved processes we adopt 
the polarized and unpolarized photonic parton distributions of Refs.~\cite{Gluck:1992zq} and~\cite{Gluck:1991jc},
respectively. In case of the polarized ones, we choose the ``maximal'' set of distributions, corresponding
to the simple assumption that the polarized and spin-averaged photonic parton distributions are equal at
some low initial scale. As shown in Ref.~\cite{Jaeger2005}, the ``minimal'' set of~\cite{Gluck:1992zq}
will lead to rather similar results, since for the kinematics we consider the process $\gamma N \rightarrow h X$
mostly probes high values of $x_\gamma$ where the inhomogeneous term in the photon evolution 
equations tends to dominate and the photonic parton distributions become relatively insensitive to the
boundary condition assumed for evolution.

The situation concerning the fragmentation functions is a bit more involved. The COMPASS data
are for {\it all} charged hadrons, specified only by charge, but not by species. In our previous paper we 
used the set of Ref.~\cite{fDSS} (``DSS07)'' which provides fragmentation functions for such 
``unidentified'' hadrons. On the other hand, fragmentation functions for pions and kaons were
substantially updated recently in~\cite{DSS_Update_2014} (``DSS14'') and~\cite{deFlorian:2017lwf} (``DSS17''), 
and it seems prudent to make use of this latest information. As pion and kaons constitute by far the largest fraction 
of produced charged hadrons, we hence also adopt the recent sets, adding pions and kaons to 
obtain an estimate for unidentified charged hadrons. We expect this approximation to be accurate
at the 90\%-level for absolute cross sections and probably even better for spin asymmetries. 
We note that COMPASS has compared their data in~\cite{Adolph:2015hta} to theoretical calculations 
that were based on the DSS14 pion fragmentation functions alone, thus neglecting heavier hadrons. 
This is still expected to catch the dominant effects. 

\subsection{Polarized and unpolarized resummed cross sections}
\begin{figure}[t]
  \centering
  \includegraphics[width=\linewidth]{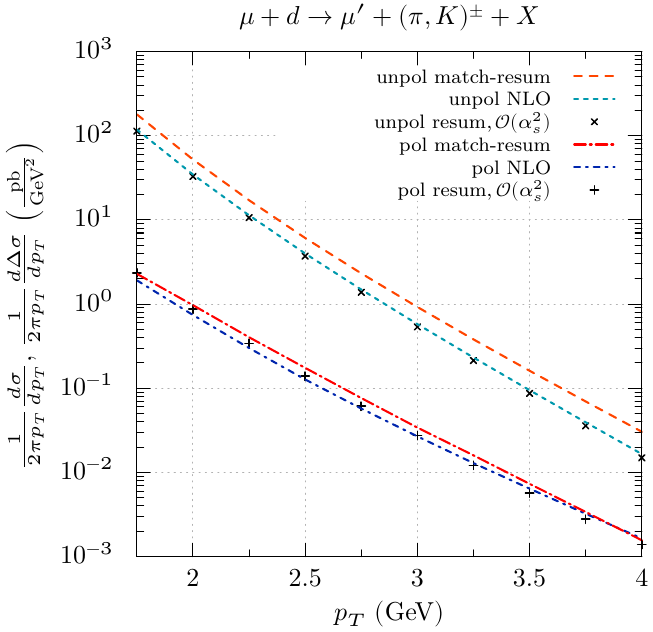}
  \caption{\label{compare_unpol-pol} {\it Spin-averaged and spin-dependent NLO and resummed (matched) 
  cross sections for combined pion and kaon production in $\mu d \rightarrow \mu' h X$
  We consider the full COMPASS rapidity range $-0.1 \leq \eta \leq 2.38$. 
  We also compare the NLO expansions of the (non-matched) resummed 
cross sections to the full NLO results (symbols)}.}
\end{figure}
In Fig.~\ref{compare_unpol-pol} we show the polarized and unpolarized cross sections for 
$\mu d \rightarrow \mu' (\pi,K)^{\pm} X$ at next-to-leading order and for the resummed case including
the matching described in Eq.~(\ref{final_matched}). The symbols in the figure show the results of 
expansions of the corresponding non-matched resummed cross sections to NLO. 
As described above, we use the DSS14 
and DSS17 fragmentation functions and sum over the contributions from produced pions and kaons.

We first consider the spin-averaged cross section. Here, the difference between the NLO cross section 
and the resummed one is sizable, especially at high $p_T$ when the threshold is more closely approached. 
The NLO expansion of the resummed cross section shows very good agreement with the full NLO result, 
illustrating that threshold resummation correctly reproduces the 
dominant parts of the cross section. This plot is very similar to the one 
shown in~\cite{MelaniesPaper} before, except for our use of the more recent fragmentation functions
(and parton distribution functions).

Concerning the polarized case, we first see that the first-order expanded resummed cross section
describes the full NLO one somewhat less accurately than in the unpolarized case (although
still rather well), indicating that subleading NLO contributions are more relevant here. The same 
observation was made in~\cite{Uebler:2015ria}. It may be partly explained in the following way: 
The direct part of the polarized cross section includes the two competing LO contributions
$\gamma q\to qg$ and $\gamma g\to q\bar{q}$, which enter with opposite signs and cancel 
to some extent. This was already observed in the NLO calculation of Ref.~\cite{Jaeger2005}.
The full spin-dependent direct resummed cross section (using DSS14 and DSS17 
fragmentation functions) is negative, while the resummed resolved contribution turns out to be 
positive and dominant for our choice of polarized parton distributions of the photon. 
On aggregate, this results in a positive polarized cross section. It is perhaps to be expected
that in the presence of such partial cancelations the expanded resummed cross section
cannot trace the full NLO one too faithfully. 

For the same reason, the polarized cross section would be expected to be quite sensitive to 
higher-order perturbative corrections. Nonetheless, threshold resummation turns out to give only
relatively small corrections to the NLO results, as may be seen from Fig.~\ref{compare_unpol-pol}. 
The resummed cross section shows only a modest enhancement over
the NLO one up to $p_T\lesssim 3.75$~GeV and even falls slightly below NLO
for yet higher $p_T$. As a consequence of this different behavior higher-order threshold effects will 
not cancel in the spin asymmetries, as we will show explicitly below. The fact that the various spin-dependent
subprocesses conspire to produce overall relatively small QCD corrections is an important outcome of 
our threshold resummation study.

\begin{figure*}[]
  \begin{center}
  \subfigure[]{\includegraphics[width=0.45\linewidth]{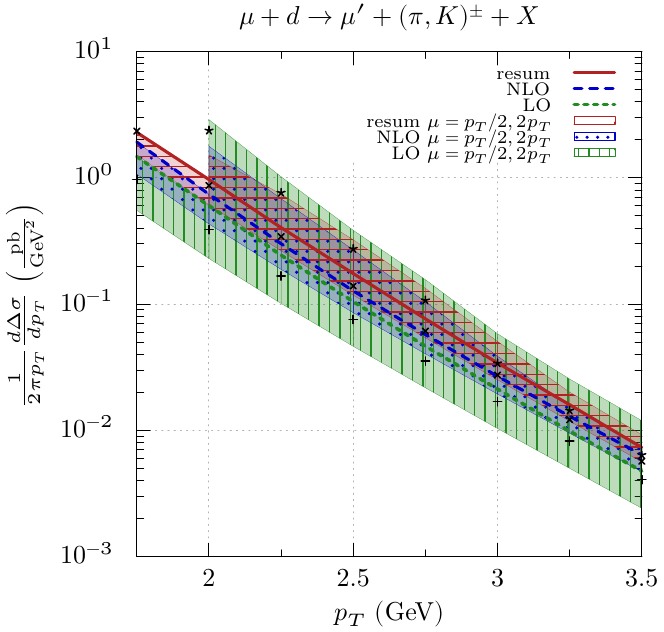}} \hspace{0.05\textwidth} 
  \subfigure[]{\includegraphics[width=0.45\linewidth]{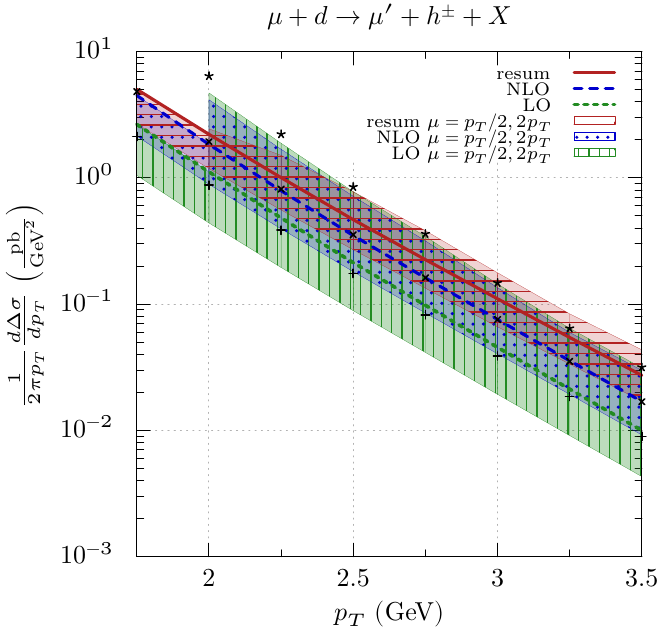}}
  \end{center}
\vspace*{-5mm}
  \caption{\label{scale-dependence} {\it (a) Scale dependence of the spin-dependent ``pion-plus-kaon'' production cross section 
in $\mu d$ scattering at LO, NLO, and for the resummed case, using the DSS14+DSS17 fragmentation functions. 
We have varied the scales $\mu= \mu_R = \mu_F= \mu_F'$ in the range $p_T/2 \leq \mu \leq 2 p_T $. 
The upper borders of the bands correspond to $\mu= p_T/2$, the lower ones to $\mu= 2 p_T$. 
We show results only when the scale exceeds 1~GeV. We use the rapidity range $-0.1 \leq \eta \leq 2.38$.
(b) Same for the DSS07 fragmentation functions.}}
\end{figure*}

In order to estimate the sensitivity of the polarized cross section to the choice of 
renormalization and factorization scales 
$\mu\equiv \mu_R = \mu_F= \mu_F'$ we vary them in the range $p_T/2 \leq \mu \leq 2 p_T $. 
The results are shown in Figs.~\ref{scale-dependence} (a) and (b) for the ``pion-plus-kaon''
fragmentation functions of DSS14 and DSS17 and the charged-hadron ones of DSS07,
respectively. In both cases we find a certain reduction of scale dependence when going
from NLO to the resummed case, although it remains unpleasantly large. On the other hand,
at LO the scale uncertainty is an order of magnitude, clearly 
demonstrating the need for higher order theory calculations.

\subsection{Double-Spin Asymmetry}
\begin{figure*}[]
  \begin{center}
  \subfigure[]{\includegraphics[width=\linewidth]{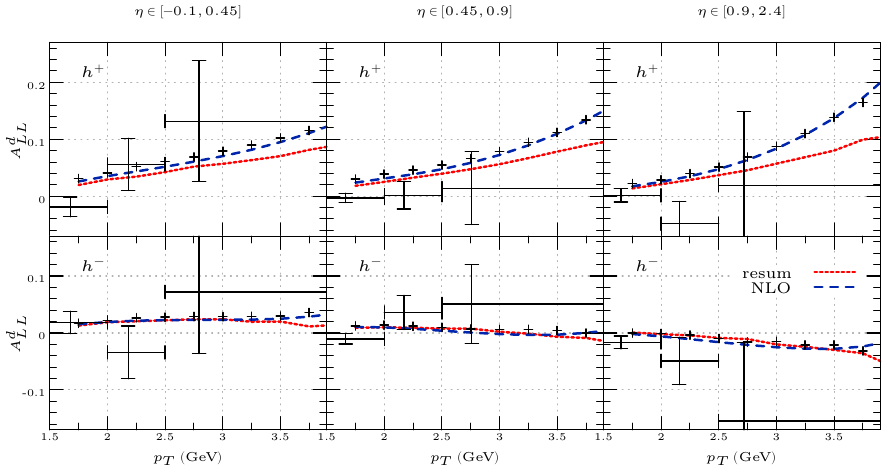}}\\
  \subfigure[]{\includegraphics[width=\linewidth]{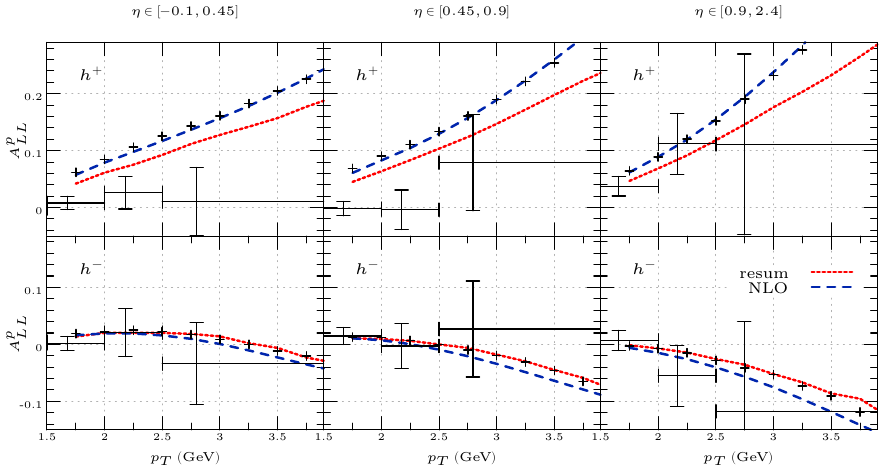}}   
  \end{center}
  \caption{\label{asymmetry_dss14+17} {\it Double-longitudinal spin asymmetries $A_{\mathrm{LL}}$ for (a) $\mu d \rightarrow \mu' h X$
   and (b) $\mu p \rightarrow \mu' h X$ in three rapidity bins, compared to the COMPASS data~\cite{Adolph:2015hta}.
   We show the NLO and resummed results using
  the combined ``pion-plus-kaon'' fragmentation functions of DSS14 and DSS17. The symbols denote the results for the asymmetry when the 
  (non-matched) resummed cross sections are expanded to first order.}}
\end{figure*}

\begin{figure*}[]
  \begin{center}
  \subfigure[]{\includegraphics[width=\linewidth]{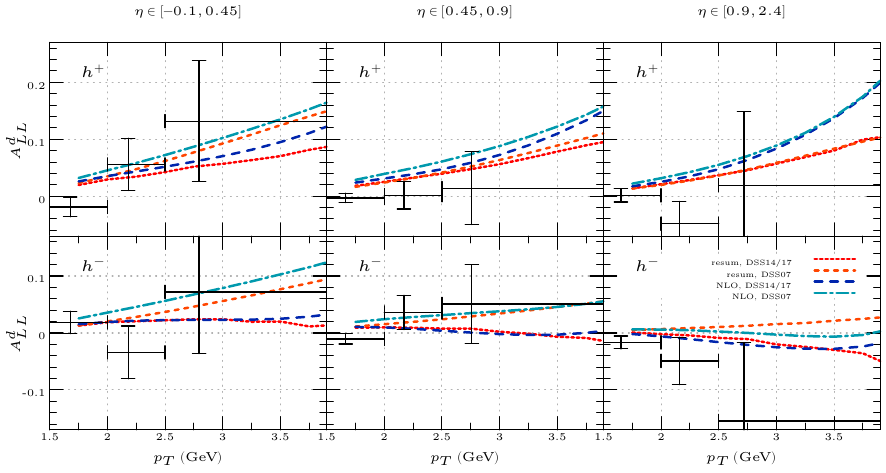}}\\ 
  \subfigure[]{\includegraphics[width=\linewidth]{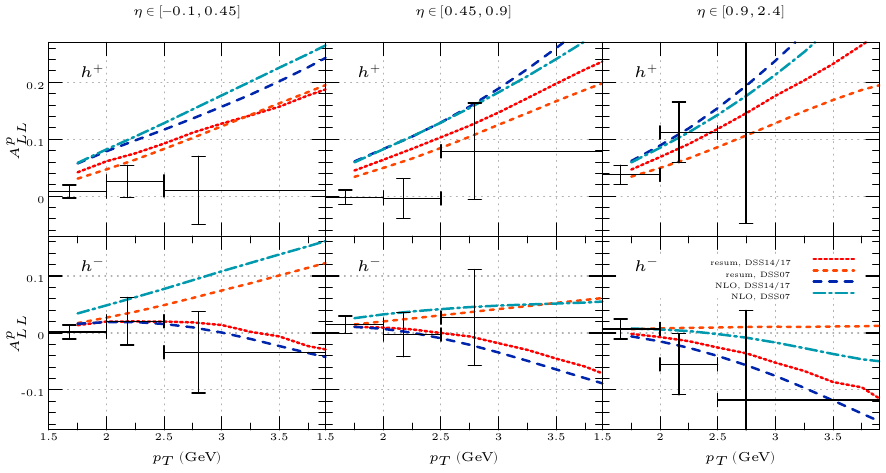}}  
  \end{center}
  \caption{\label{asymmetry_compareDSS} {\it Same as Fig.~\ref{asymmetry_dss14+17},
  but now also showing the results for the DSS07 set of charged-hadron fragmentation functions.}}
\end{figure*}

Our results for the double longitudinal spin asymmetries $A_{\mathrm{LL}}$ for single-inclusive charged
hadron production with a deuteron or a proton target are shown in Figs.~\ref{asymmetry_dss14+17}
and~\ref{asymmetry_compareDSS}, compared to the COMPASS data~\cite{Adolph:2015hta}.
The asymmetry is defined as the ratio of the spin-dependent and the spin-averaged cross section:
\begin{align}
 A_{\mathrm{LL}}=\frac{d\Delta \sigma}{d \sigma} \,.
\end{align}
Figure~\ref{asymmetry_dss14+17} shows our NLO and resummed results for the ``pion-plus-kaon''
fragmentation functions of DSS14~\cite{DSS_Update_2014} and DSS17~\cite{deFlorian:2017lwf}. 
We investigate the three rapidity bins $[-0.1,0.45]$, $[0.45,0.9]$ and $[0.9,2.4]$. 
The symbols in the figure show the results for the asymmetry when the (non-matched) resummed 
polarized and spin-averaged cross sections are expanded to first order.

As was to be anticipated from the results shown in Fig.~\ref{compare_unpol-pol}, the 
threshold effects do not cancel in the double-longitudinal spin asymmetry $A_{\mathrm{LL}}$ and
rather tend to decrease the asymmetry when going from NLO to the resummed case. 
This implies that threshold logarithms beyond NLO cannot be ignored and have to be 
resummed to all orders. 

Within the rather large experimental uncertainties we find an overall fair agreement
between our resummed results and the COMPASS data. Some tension is observed perhaps
for positively charged hadrons produced off a proton target. Resummation, especially that
for the resolved contribution, tends to improve the description of the experimental results.

Figure~\ref{asymmetry_compareDSS} investigates the sensitivity to the choice of fragmentation functions,
by comparing the results shown in Fig.~\ref{asymmetry_dss14+17} with those obtained for
the DSS07 set~\cite{fDSS}. We find that the more recent fragmentation functions lead to a significantly
better agreement with the experimental data, especially for the case of negatively charged hadrons
produced off a proton target. This improvement is the result of an interplay of several features.
By and large, the cross section for $\mu p \rightarrow \mu' h^- X$
is expected to be more sensitive to ``non-favored'' fragmentation 
functions than the one for production of positive hadrons. These functions are now a little better determined
from data taken in semi-inclusive deep inelastic scattering. Furthermore, as mentioned earlier,
in the spin-dependent case there are several competing contributions (of partly opposite signs) 
to the cross section. As a result, even relatively small differences in the fragmentation functions
can make a sizable effect. Even the gluon fragmentation, which is smaller in DSS14 than in DSS07,
plays a role here. These findings clearly demonstrate the need for further improved determinations
of the fragmentation functions. 

%
%
%
\section{Conclusions and outlook \label{sec5}}
We have presented a detailed phenomenological study of the impact of next-to-leading logarithmic 
threshold resummation on the spin-dependent high-$p_T$ photoproduction cross section
for $\mu N \rightarrow \mu' h X$ and on the corresponding double-longitudinal spin asymmetry $A_{\mathrm{LL}}$. 
Compared to our previous calculation~\cite{Uebler:2015ria} which focused on the direct ``point-like'' contribution to 
the cross section, we have now also included resummation for the resolved-photon contribution, 
for which the photon behaves like a hadron.

Our phenomenological studies have shown that for the kinematics relevant for the COMPASS experiment
the spin-dependent cross section receives smaller corrections from resummation than the 
spin-averaged one. As a result, the threshold corrections do not cancel in 
the double-spin asymmetry but rather tend to decrease the asymmetry, leading to an overall
better agreement between the COMPASS data and theory. Thus, inclusion of threshold resummation
is vital for phenomenology for COMPASS kinematics. We expect this to remain true
also at a future Electron Ion Collider (EIC) where the process $\mu N \rightarrow \mu' h X$ 
could be explored with unprecedented precision and kinematic reach~\cite{elke}.
While our present calculation marks the state of the art for theoretical
studies of $\mu N \rightarrow \mu' h X$ it appears that at an EIC an even 
higher level of theoretical precision will be needed. This may be achieved, for example, 
by extending our resummation studies to next-to-next-to-leading logarithmic accuracy.

We have also found that the parton-to-hadron fragmentation functions have a strong
impact on the size and shape of the predicted spin asymmetries $A_{\mathrm{LL}}$.
The most recent sets which contain much more up-to-date experimental information
help to improve the description of the COMPASS data. Nevertheless, the 
fragmentation functions arguably remain a primary source of systematic 
theoretical uncertainty, so that continued improvements of the functions are
necessary. Promising avenues in this direction are perhaps offered by studies
of hadron fragmentation inside jets~\cite{Kaufmann:2015hma,Chien:2015ctp,Anderle:2017cgl,procura}. 

%
%
\begin{acknowledgments}
We are grateful to Y. Bedfer, F. Kunne, M. Levillain, C. Mar\-chand, M. Pfeuffer, G. Sterman, and M. Stratmann for
helpful discussions. This work was supported in part by 
the ``Bundesministerium f\"{u}r Bildung und Forschung'' (BMBF) grants no. 05P15WRCCA,
05P12VTCTG and 05P15VTCA1.
\end{acknowledgments}
%
%
%
%
%

%
%
\end{document}